\documentclass{article}

\PassOptionsToPackage{numbers, square, sort}{natbib}

\usepackage[final]{neurips_2019}

     \usepackage[final]{neurips_2019}

\usepackage[utf8]{inputenc} 
\usepackage[T1]{fontenc}    
\usepackage[hidelinks]{hyperref}       
\usepackage{url}            
\usepackage{booktabs}       
\usepackage{tabularx}
\usepackage{amsfonts}       
\usepackage{nicefrac}       
\usepackage{microtype}      
\usepackage{amsmath}
\usepackage{xcolor}
\usepackage{siunitx}
\usepackage{tikz}
\usepackage[normalem]{ulem}

\usepackage{enumitem}
\setlist[description]{leftmargin=0pt,labelindent=0pt}

\usetikzlibrary{arrows.meta}
\newcommand{\niceArrow}{-{Latex[length=2.1mm, width=1.5mm]}}

\title{A Data Scientist's Guide to Streamflow Prediction}

\author{%
  Martin Gauch\thanks{Corresponding author: Martin Gauch, \texttt{martin.gauch@uwaterloo.ca}} ~ and Jimmy Lin\vspace{0.1cm}\\
  David R.\ Cheriton School of Computer Science\\University of Waterloo, ON, Canada
}

\definecolor{jmcolor}{RGB}{255,127,0}

\newcommand{\ignore}[1]{}

\begin{document}

\maketitle

\begin{abstract}
In recent years, the paradigms of data-driven science have become essential components of physical sciences, particularly in geophysical disciplines such as climatology.
The field of hydrology is one of these disciplines where machine learning and data-driven models have attracted significant attention.
This offers significant potential for data scientists' contributions to hydrologic research.
As in every interdisciplinary research effort, an initial mutual understanding of the domain is key to successful work later on.
In this work, we focus on the element of hydrologic rainfall--runoff models and their application to forecast floods and predict streamflow, the volume of water flowing in a river.
This guide aims to help interested data scientists gain an understanding of the problem, the hydrologic concepts involved, and the details that come up along the way.
We have captured lessons that we have learned while ``coming up to speed'' on streamflow prediction and hope that our experiences will be useful to the community.
\end{abstract}

\section{The Problem in a Nutshell}

At its core, streamflow prediction is a spatio-temporal prediction effort that tries to predict the amount of water flowing past a given point of a river, given the past measurements of river streamflow, past meteorological variables known as forcings (e.g., air temperature, humidity, amount of precipitation), and geophysical variables (e.g., land cover, soil, elevation).

Thus, models used for this purpose take as input meteorological forcings and geophysical information and output a sequence of streamflow predictions at some temporal granularity at some location.
This temporal granularity can range anywhere from hours to years: During flood events, we might need models at an hourly scale to predict exact peak times, whereas climate change analyses focus on changes over several years.

Under the influence of climate change, both floods and droughts are likely to become increasingly common and pose huge dangers on humans, flora, and fauna~\cite{IPCC2012, Trenberth2014drought}.
During these extreme events, accurate streamflow predictions are an important prerequisite for mitigative action, as they allow authorities to proactively warn citizens, manage water regulation facilities, and direct help to where it will be needed the most.

\section{A Lot More Nuance}

\subsection{Understanding \textit{What} We're Actually Predicting}

\begin{figure}[t]
    \centering
    \includegraphics[width=0.8\textwidth]{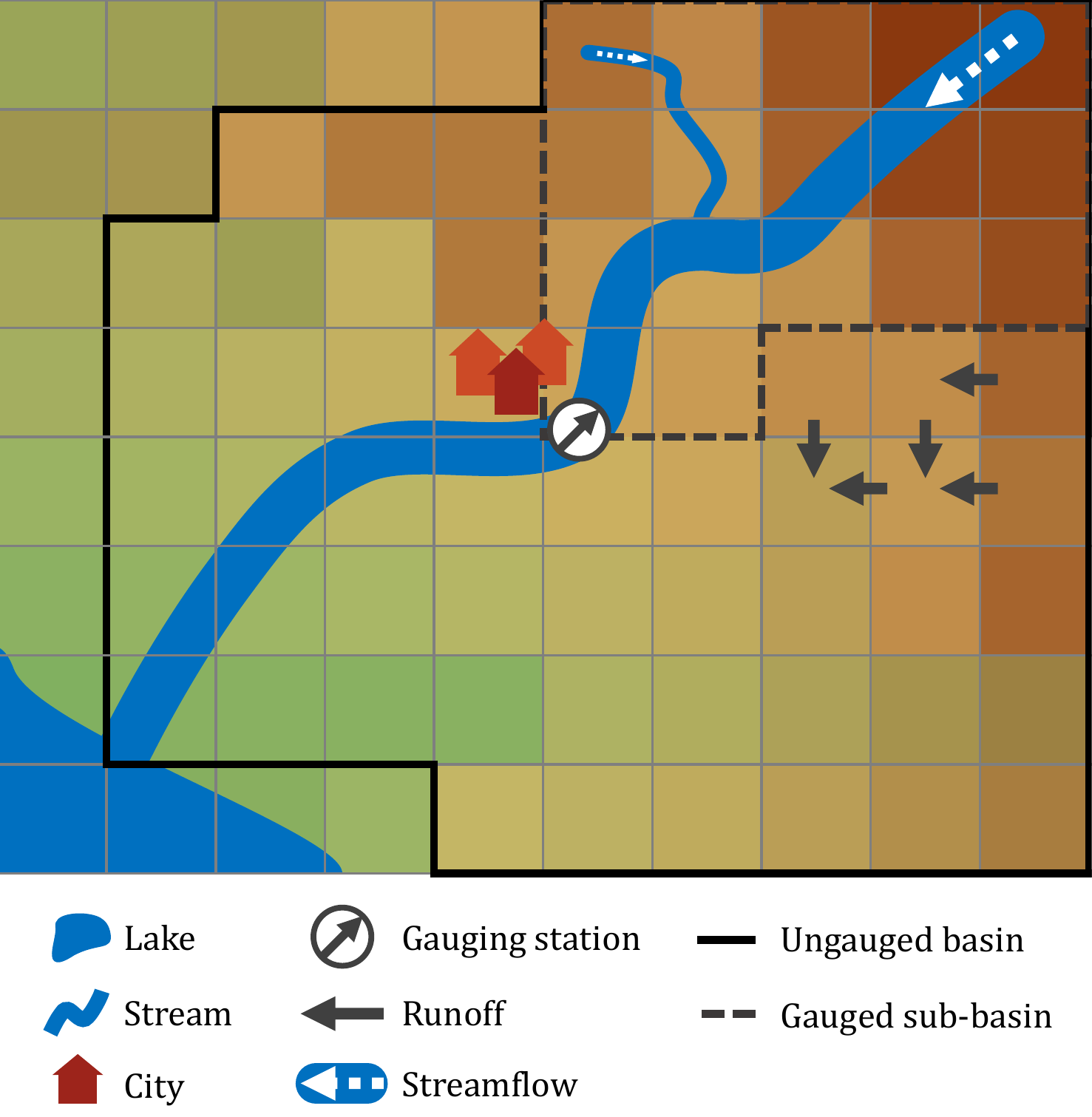}
    \caption{Illustration of streamflow, runoff, basins, and sub-basins. The background gradient represents elevation (brown = high, green = low). From the elevation, we can derive the outline for the ungauged basin (solid black line). The highlighted sub-basin is gauged near a city (dashed line). Unlike streamflow, runoff has a meaning at every grid cell, and the cells are connected through downstream-relationships (gray arrows as a few examples).}
    \label{fig:flow}
\end{figure}

Usually, hydrologists model streamflow at geographically prominent points such as a city, the mouth of a river, or the outflow to a lake. 
This point is referred to as the \textit{outlet}. Given an outlet point, we can delineate the \textit{upstream area} from which all water drains towards the outlet.
This upstream area for the outlet point has several synonymous names: \textit{basin}, \textit{catchment}, or \textit{watershed}.
(As a side note, the term \textit{watershed} may also refer to a \textit{drainage divide}: the border between two neighboring catchments, where precipitation divides into water flowing in one or the other catchment.)
We call a catchment a \textit{gauged basin} if its outlet is equipped with a gauging station to measure the time series of actual streamflow (known as \textit{hydrograph}).
Catchments without gauging station are called \textit{ungauged basins}.
Mostly, the delineation of catchments is based on topological information, a natural implication of water following downward gradients.
Clearly, the size of a basin depends on the selected outlet.
The further upstream along a river the outlet, the smaller the basin.
In addition, rivers and lakes drain into each other.
Consequently, basins contain nested \textit{sub-basins} (or \textit{sub-catchments} or \textit{sub-watersheds}).
The example in \autoref{fig:flow} illustrates these concepts: The solid black line outlines an ungauged basin, whereas the gauge station at the city represents the outlet of a gauged sub-basin (dashed gray line).
More formally, we can think of the river network as a directed graph where each node is a sub-basin or grid cell and each edge represents a downstream-relationship.
For instance, an edge from node $A$ to $B$ indicates that all water flowing out of cell $A$ drains into cell $B$.
The gray arrows in \autoref{fig:flow} show a few example downstream relations between grid cells---note how the arrows point to the adjacent cell with the lowest elevation (i.e., with the greenest hue).

Hydrologists distinguish two concepts: \textit{runoff} and \textit{streamflow}.
Runoff refers to an amount of water that flows inside or between basins and their building blocks such as hillslope.
This can happen on the surface (surface runoff), near the surface (interflow), or through the groundwater (groundwater seepage).
Commonly, the runoff is expressed in millimetres during a fixed time period such as one day.
An intuitive explanation for this notion is the hypothetical depth of water if all outflowing water was equally distributed over the geographical area of interested (e.g., a grid cell, hill slope, or basin).

Streamflow on the other hand is the amount of water that flows through the cross-section at a point along a stream and usually expressed as cubic meters per second.
We can think of streamflow as the aggregation of runoff from across the whole basin that drains into a river.
At any point in a river, hydrologists can set up a gauging station that measures streamflow.
An often neglected yet important detail of these measurements is that they come with an uncertainty.
Depending on the gauging technique, the measurements can have errors of \SI{5}{\percent} or more.
This error can increase during a flood event and in the worst case, the flood can sweep away the whole gauging device~\cite{Beven2012Primer}.

Note that runoff is defined for arbitrary areas, even if there is no stream in the area.
Hence, we can talk about the runoff in or from an area; this is the amount of water that the area contributes to the streamflow.
Streamflow, meanwhile, is only defined at points on a stream or river, so we cannot talk about the streamflow at arbitrary points in space.

While it is relatively easy to measure streamflow at any point along a stream, runoff is much harder to measure---especially for large basins.
Hence, models are often evaluated against streamflow rather than runoff.
This is also why streamflow modeling is commonly called rainfall--runoff modeling:\ internally, many models predict runoff, but the evaluation is based on streamflow.
To obtain streamflow from runoff values, hydrologists use \textit{routing models}.
A routing model is based on the directed graph of downstream-relations in the river network.
During routing, we multiply the runoff in \si{\milli\metre\per\day} by the area (including unit conversion to \si{\metre\per\day} and \si{\square\metre}), which yields the total water volume in \si{\cubic\metre} per day.
Next, we accumulate the runoff along the edges of the routing graph to obtain daily streamflow.
To comply with SI base units, we finally normalize the streamflow to per-second values in \si{\cubic\metre\per\second}.

Hence, in a nutshell, streamflow is routed runoff at a point along a stream.

\subsection{Understanding the Types of Input Data}
\label{section:input-data}

There are two broad types of input data:\ forcings and geophysical data.
Forcings are time series of meteorological data such as precipitation or temperature.
Their name stems from the fact that these data are needed to run, or \textit{force}, the model.
Geophysical data are static (non-time-series) data that provide information such as land cover, soil, or elevation.
Clearly, these variables are not \textit{actually} static; land usage, for example, changes over time---but generally at a much slower pace than meteorological information.
In most cases other than long-term simulations (for example, multi-decade climate change analyses), such data can be safely treated as static.

If we forecast into the future, we have no choice but to rely on forcings that are predictions themselves, for instance, from a numerical weather prediction model.
For simulations of historical streamflow, the forcings can, if available, be observed data.
Since observations are generally sparse, however, we commonly use forcings that are generated by other models, for instance, climate reanalyses.
Climate reanalyses are a type of data product that use one consistent parameterization to simulate long periods of historical meteorologic variables.
They use data assimilation techniques to ensure that the output is as close to the available observations as possible.
What makes these reanalysis products valuable compared to short-term weather predictions is their consistency:\ environmental agencies update their weather prediction models frequently, which leads to inconsistent variable contents over time.

The data a model takes as input can come in two forms, depending on whether the model is lumped, semi-distributed, or fully-distributed.
Lumped models take forcings and geophysical data that are aggregated to a basin-level as input.
A lumped model that predicts streamflow for a certain basin will need ``pointwise'' forcing time series and geophysical data.
For example, a lumped model of the gauged sub-basin in \autoref{fig:flow} might take as input the sub-basin's minimum and maximum temperature, the cumulative precipitation, and the mean elevation calculated across all cells within the dashed sub-basin outline.

Most publicly available streamflow datasets are well-suited for lumped models, as there is a long history of streamflow recordings through gauging stations along rivers, where scientists have been measuring streamflow and meteorological variables for decades.
One common collection of such datasets are the \textit{Catchment Attributes and Meteorology for Large-sample Studies} (CAMELS) that contain data for basins across the continental United States~\cite{Newman2014Camels, Addor2017Camels}.
Recently, similar datasets for basins in Chile, CAMELS-CL~\cite{Alvarez2018CamelsCL}, and Great Britain, CAMELS-GB~\cite{Coxon2020CamelsGB}, have been published, and datasets for other regions are in development.
It is important to note, however, that each of these datasets is based on its own meteorological forcing product.
Consequently, the resulting lumped datasets have different biases and we cannot simply apply a model that was, for example, trained on the U.S.-CAMELS to basins from the Chilean version.

Distributed and semi-distributed models take a spatial grid of input variables, although semi-distributed models internally aggregate the grid cells into \textit{response units}, as we will see in \autoref{sec:model-types}.
For example, a distributed model of the gauged sub-basin in \autoref{fig:flow} would take as input forcings and geophysical data individually for each cell within the dashed sub-basin outline.

Such gridded forcings for large spatial extents are much harder to get hold of.
As it is impossible to actually measure the meteorological variables in all grid cells, these datasets require either computationally expensive simulations that spatially interpolate between the available measurements, or data acquired through remote sensing, which is costly and not accurately possible for all variables of interest.
To make things worse, the different available measurements will likely need preprocessing to be used within one dataset, as they might be in different units, use different measurement techniques, or have other inconsistent characteristics.

\subsection{Understanding \textit{How} We're Actually Predicting}
\label{sec:model-types}

\subsubsection{Spectrum of Models}
There are various mutually orthogonal ways to classify hydrologic rainfall--runoff models.
First, we distinguish between \textit{process-based models} that explicitly model hydrologic processes and \textit{data-driven models} that encompass machine learning models.
\begin{description}
    \item[Process-Based Models] --- Process-based rainfall--runoff models are simplified representations of the underlying dominant natural processes that generate runoff or streamflow from rainfall.
    The hydrologic processes involved are, for example, evaporation from the surface or transpiration from plants (both of which depend on solar radiation), or soil infiltration (which depends on the soil's capacity of holding water; different soil types have different water capacities).
    Process-based models are also sometimes called physically-based models~\cite{Beven1989Physically}, which is somewhat misleading, since these models model simplified abstractions rather than the actual physical processes (in fact, the actual processes are still partly unknown).

    Process-based models usually consist of \textit{model states} and \textit{model parameters} that govern the behavior of model states.
    Model parameters are the ``knobs'' that are adjusted during the training phase (which hydrologists call the \textit{calibration} phase).
    Data scientists may compare these parameters to the weights in a neural network (note, this is not to say that hydrologic models are neural networks, but only to illustrate the kind of parameter).
    Model states (or \textit{storages}) meanwhile are ``memory cells'' that have an initial value, and are updated by the model equations, depending on the input.
    Such states can, for example, simulate current snow depth or soil moisture.
    Sub-zero temperature, combined with precipitation, will increase the value of the snow depth state. This snow will eventually melt, so the model needs the state as a memory of snow events.
    In our illustrative comparison, these states are similar to memory states in a recurrent neural network, with the difference that hydrologists have a physical interpretation of a state's value.

    To provide good predictions after calibration, it is key that the initial model states are adjusted to the beginning of the test period.
    For instance, an initial snow depth state of zero might not be suitable for a test period in winter.
    To ensure good initial states, as it usually hard to estimate the initial states, hydrologists prepend \textit{warm-up periods} before the training and test periods.
    During warm-up, the model takes input data and updates its states based on the input, but the period is not considered in evaluation metrics.
    As the test period commonly immediately succeeds the training period, there is often no need for explicit warm-up before testing.
    The model simply runs through the training warm-up and training periods before it starts to generate test predictions.
    Refer to \autoref{sec:abcmodel} for an example of a simple process-based model.

    \item[Data-Driven Models] --- Data-driven models on the other hand are what is familiar to data scientists: machine-learned models that do not try to model any physical dependencies, but just generate a prediction based on historical training data.
    Since we try to predict a continuous time series, we generally use regression models.
    A common architecture for time-series modeling are Long Short-Term Memory networks (LSTMs), but we can also apply models that are not directly tailored to time-series prediction such as regression tree-based models~\cite{Kratzert2019Benchmark, Gauch2019Feeding}.
    In this case, we simply feed the model a fixed-history window of forcings to predict one time step of streamflow.
    Of course, one can image hybrid models that blur the lines between process-based and data-driven approaches---a research direction known as \textit{theory-guided data science} or \textit{physics-guided machine learning}~\cite{Karpatne2017Theory, Jia2020Physics}.
    For instance, we might train a data-driven regression model to perform error correction on a process-based model, and obtain a hybrid ensemble model.
\end{description}

Another way to classify streamflow models is grounded in their degree of spatial distribution.
Models can be either \textit{lumped}, \textit{semi-distributed}, or \textit{distributed}.
\begin{description}
   
    \item[Lumped Models] --- Lumped models do not operate on gridded data at all.
    Instead, they require aggregated, or lumped, forcings for one point in space as input, and provide a prediction for this point in space.
    Hence, if our forcings are spatially distributed, we need to aggregate the variables across the whole basin before passing them to the model.
    An example for a process-based lumped model is the \textit{Large Basin Runoff Model} (LBRM), used in production for the U.S.\ Army Corps of Engineers' (USACE) water level predictions in the Great Lakes region~\cite{Croley1983LBRM}.
    Up to now, most purely data-driven models are also lumped setups~\cite{Kratzert2019Benchmark, Gauch2019Feeding}.
    The reason for this is that a lumped model's design is much more straight-forward, since it does not have to deal with differently-sized and -shaped basins and the resulting varying input dimensions.
    
    \item[Distributed and Semi-Distributed Models] --- Distributed models are computationally the most expensive, as they operate on gridded forcings and output runoff for each grid cell.
    They require meteorological forcings and potentially geophysical input data for an entire area.
    
    Semi-distributed models fall conceptually between lumped and fully-distributed models.
    These models operate on arbitrarily-shaped polygonal spatial extents, sometimes called \textit{hydrologic response units}, or HRUs.
    The model assumes the area within each unit to have identical hydrologic properties.
    HRUs are usually derived from the overlay of various geophysical maps, such as soil, vegetation, and slope characteristics.
    Each continuous parcel of identical properties forms a response unit.
    Although semi-distributed models might take fully-distributed forcing grids as input, they will internally process them per HRU and, for instance, only have one set of parameters per HRU rather than one per grid cell.
    Hence, semi-distributed models are in some sense lumped models at the scale of HRUs rather than the scale of entire catchments.
    As the HRUs will usually be larger than grid cells in distributed models, semi-distributed offer a compromise between computational efficiency and prediction accuracy.
    One example for a semi-distributed model is the \textit{Variable Infiltration Capacity} model (VIC)~\cite{Liang1994vic, Hamman2018vic}.
\end{description}

\subsubsection{Model Transferability}
Although a data scientist might speak generically about models making predictions, hydrologist are careful to distinguish between distinct types of predictions, depending on the temporal and spatial domains to which we apply a model:
\begin{description}
    \item[Transferability in Time: Simulation, Forecasting, and Hindcasting Models] --- Simulation models predict \textit{past} streamflow---they try to reproduce past system states. 
    Forecasting models predict streamflow for \textit{future} dates.
    Of course, it is entirely possible to act as if the input were only available up to a certain time step in the past, and then ``forecast'' the past streamflow after this time step; this process is known as \textit{hindcasting}.
    The word \textit{prediction} can, meanwhile, refer to any of these classes.
    We refer to Beven and Young~\cite{Beven2013Semantics} for even more fine-grained distinctions between different prediction models in the hydrologic literature.
    
    \item[Transferability in Space: Modeling Across Locations] ---
    Many models, especially lumped setups, are designed to be trained and evaluated on one single basin or a fixed set of basins.
    This means that we cannot expect reasonable predictions when we evaluate the model on basins that were not part of the training procedure, a process known as \textit{spatial validation}.
    Instead, these models are only evaluated in \textit{temporal validation}, where we use the same basins as in training, but a different time period.
    Clearly, achieving good temporal validation results is much easier than achieving good spatial validation results, as the relationships between forcings and streamflow remain relatively static over time, while the geophysical properties of other basins can make for completely different relationships (although climate change threatens to alter these ``static'' relations, which may make temporal validation harder in the future).
    More sophisticated models try to adapt their streamflow generation to the geophysical properties of a basin to achieve some degree of spatial generalization.
    Further, there exist lots of so-called \textit{regionalization} techniques to transfer a model to a new set of basins, for instance, by scaling the predictions according to the basin area.

\end{description}

\subsection{Performance Metrics}
\label{sec:evaluation}
Given time series of streamflow ground truth $\mathbf{y}$ and predictions $\mathbf{\hat{y}}$, there are a variety of metrics to quantify the error.
A very common metric is the \textit{Nash--Sutcliffe model efficiency coefficient} (NSE):
\begin{equation}
\label{eq:nse}
\begin{aligned}
    \text{NSE} &= 1 - \frac{\sum\limits_{t=1}^T{(\hat{y}_t - y_t)^2}}{\sum\limits_{t=1}^T{(y_t - \bar{y})^2}} \\
    &= 1- \frac{\text{MSE}}{\frac{1}{T}\sum\limits_{t=1}^T{(y_t - \bar{y})^2}}
\end{aligned}
\end{equation}
$\bar{y}$ denotes the mean observed streamflow, hence, the denominator is the variance of the streamflow observations $\mathbf{y}$. 
\autoref{eq:nse} shows that NSE is strongly correlated with the mean squared error (MSE), which will be more familiar to data scientists~\cite{Gupta2009NSE}.
NSE values range from $-\infty$ to $1$, where $1$ corresponds to perfect predictions and $0$ corresponds to a model that constantly predicts the station's mean ground truth.

Another common metric is the \textit{Percent Bias} (PBIAS), which measures the tendency of predictions to be larger or smaller than the ground truth:
\begin{align}
    \label{eq:pbias}
    \text{PBIAS} = 100 \times \frac{\sum\limits_{t=1}^T{(\hat{y}_t - y_t)}}{\sum\limits_{t=1}^T{y_t}}
\end{align}
PBIAS values range between $-\infty$ and $\infty$, where the best value is zero.
If a model's PBIAS is larger than zero, this means that it over-estimates streamflow on average.
Conversely, a negative PBIAS corresponds to models that generally under-estimate streamflow.

Related to PBIAS is the concept of \textit{water balance}.
Process-based models are inherently designed in such a way that they conserve mass:\ they do not generate water out of nowhere.
In other words, for any interval of time, the precipitation over a basin equals the sum of evapotranspiration (water that evaporates from the surface or transpiration from plants), streamflow, and changes in water storage (note that for the sake of simplicity, we assume a closed system where adjacent basins do not exchange any water, e.g., through groundwater in- and outflow).

\section{Formal Notation for the Problem}
\begin{table}[t]
    \centering
    \caption{Notation}
    \label{tab:notation}
    \begin{tabularx}{0.9\textwidth}{lp{2.5cm}X}
        \toprule
        Symbol & Range & Use \\
        \midrule
        $t, T$ & $\mathbb{R}$ & A specific time step, total number of time steps \\
        $\mathbf{y}, \mathbf{\hat{y}}$ & $\mathbb{R}^T$ & Actual and predicted streamflow \\
        $d$ & $\mathbb{N}_+$ & Number of forcing variables \\
        $h, w$ & $\mathbb{N}_+$ & Size of a grid of forcings. (Note that this is a simplification to mathematically express the dimensions. In reality, we can have grids of arbitrary shape and size---just as basins have different shapes.) \\
        $\sigma$ & $\mathbb{N}_0$ & Number of static variables \\
        $X$ & $\mathbb{R}^{d \times T}$ or $\mathbb{R}^{(d \times (h \times w)) \times T}$ & Forcing variables for lumped and semi-distributed or distributed models\\
        $\Sigma$ & $\mathbb{R}^{\sigma}$ or $\mathbb{R}^{\sigma \times (h \times w)}$ & Static, non-time-series model input for lumped and semi-distributed or distributed models\\
        \bottomrule
    \end{tabularx}
\end{table}

We denote the hydrograph for one gauging station, measured in \si{\cubic\metre\per\second}, as
\begin{equation}
    \mathbf{y} = (y_1, \dots, y_T) \in \mathbb{R}^T
\end{equation}
Note that hydrologists commonly refer to streamflow as $\mathbf{q}$ or $\mathbf{Q}$.
We use $\mathbf{y}$ to align with the usual notation for dependent variables in the machine learning literature, and denote predictions as $\mathbf{\hat{y}}$.
Note that the time steps can arbitrarily granular, but typical step sizes are hourly, daily, monthly, or yearly.

The most important input variables for streamflow prediction are meteorological time-series forcings such as air temperature, precipitation, wind speed, etc., which we denote with $X$.
In addition to the time-series forcings $X$, we denote the geophysical information that is assumed not to change over time with $\Sigma$.
Examples for static input data include digital elevation, land cover, and soil maps.
The exact format of $X$ and $\Sigma$ depends on whether the model is lumped or (semi-)distributed.
For lumped models, the input data are aggregated to one single basin-wide time series for each of the $\sigma$ geophysical and the $d$ forcing variables:
\begin{equation}
\begin{aligned}
    \Sigma &\in \mathbb{R}^\sigma,\\
    X &= (X_1, \dots, X_T)\\
      &= \left( \begin{bmatrix} x_1^1 \\ \vdots \\ x_1^d \end{bmatrix},
          \dots,
          \begin{bmatrix} x_T^1 \\ \vdots \\ x_T^d \end{bmatrix} \right)
          \in \mathbb{R}^{d\times T}
\end{aligned}
\end{equation}

For (semi-)distributed models, the input data are a spatially-distributed grid of cells for each geophysical or meteorological variable. The shape of these grids is a discretization of the basin shape, but for simplicity we denote it as $h \times w$ grid cells:
\begin{equation}
\begin{aligned}
    \Sigma &\in \mathbb{R}^{\sigma \times (h \times w)},\\
    X &= (X_1, \dots, X_T), \\
    \text{where } X_t &= \begin{bmatrix} 
                \mathbf{x}_t^{1,1} & \dots & \mathbf{x}_t^{1,w} \\
                \vdots & \ddots & \vdots \\
                \mathbf{x}_t^{h,1} & \dots & \mathbf{x}_t^{h,w} 
            \end{bmatrix} \in \mathbb{R}^{d \times (h \times w)}, \\
    \text{where } \mathbf{x}_t^{i,j} &= \begin{bmatrix} x_{t,1}^{i,j} \\ \vdots \\ x_{t,d}^{i,j} \end{bmatrix} \in \mathbb{R}^d
\end{aligned}
\end{equation}

In the standard setup with gauged basins, along with $\Sigma$ we're given $\mathcal{D} = \{(X_t, y_t)\}_{t=1}^{T}$ as training data, which directly aligns with the standard supervised machine learning setup. I.e., our task is to learn the parameters $\theta$ of a model $f$ that minimize some loss or maximize some metric (see \autoref{sec:evaluation}):
\begin{equation}
\label{eq:model}
f(\Sigma, X_{t-k+1}, \dots, X_{t}; \theta) = \hat{y}_t
\end{equation}

It is important to note that the model $f$ \textit{does not} take past streamflow as input. 
This may seem unreasonable from a mere prediction-quality viewpoint, since streamflow usually has a high auto-correlation.
In hydrology, however, hydrologists are not only interested in obtaining good predictions, but also in explaining the physical drivers of streamflow.
Clearly, a model that requires past streamflow to provide good predictions does not fully capture the physical processes behind streamflow: Water that has already passed a gauging station can no longer affect that station's streamflow.
Further, we might want to apply the model to ungauged basins, where simply no measurements exist that could be used as input.

In many operational forecasting scenarios, however, we care much more about accurate predictions than hydrologic understanding.
In these cases it does make sense to explicitly use previous time steps of streamflow to generate predictions.
This procedure is called \textit{data assimilation}:\ models explicitly use streamflow values $y_{t-k+1}, \dots, y_{t-1}$ to predict streamflow $\hat{y}_t$ (or, when forecasting the future, they use $\hat{y}_{t-k+1}, \dots, \hat{y}_{t-1}$).
Process-based models that perform data assimilation usually do so by changing internal model states, while model parameters remain fixed after the initial calibration.
Forecasting without assimilation is called \textit{open-loop forecasting}.

The above problem setup is for a single gauge, but ideally, we would strive for a model that applies universally to \textit{any} basin---even if it does not contain a gauging station.
This motivates the case of ungauged basins, where we don't have the ground truth $\mathbf{y}$ for (some of) the basins that we're predicting.
That is, we are attempting to learn models from one location and apply it to another---the location can be nearby (e.g., in the same watershed) or halfway around the world.

\section*{Acknowledgments}
The authors thank Shervan Gharari from the University of Saskatchewan for his valuable comments and suggestions throughout the creation of this paper.
We further thank the Global Water Futures program for their financial support. 
This research was undertaken thanks in part to funding from the Canada First Research Excellence Fund.

\bibliographystyle{abbrvnat}
\bibliography{references}

\begin{thebibliography}{18}
\providecommand{\natexlab}[1]{#1}
\providecommand{\url}[1]{\texttt{#1}}
\expandafter\ifx\csname urlstyle\endcsname\relax
  \providecommand{\doi}[1]{doi: #1}\else
  \providecommand{\doi}{doi: \begingroup \urlstyle{rm}\Url}\fi

\bibitem[Addor et~al.(2017)Addor, Newman, Mizukami, and Clark]{Addor2017Camels}
N.~Addor, A.~Newman, N.~Mizukami, and M.~P. Clark.
\newblock The {CAMELS} data set: catchment attributes and meteorology for
  large-sample studies.
\newblock \emph{Hydrology and Earth System Sciences}, 21\penalty0
  (10):\penalty0 5293--5313, 2017.

\bibitem[Alvarez-Garreton et~al.(2018)Alvarez-Garreton, Mendoza, Boisier,
  Addor, Galleguillos, Zambrano-Bigiarini, Lara, Puelma, Cortes, Garreaud,
  McPhee, and Ayala]{Alvarez2018CamelsCL}
C.~Alvarez-Garreton, P.~A. Mendoza, J.~P. Boisier, N.~Addor, M.~Galleguillos,
  M.~Zambrano-Bigiarini, A.~Lara, C.~Puelma, G.~Cortes, R.~Garreaud, J.~McPhee,
  and A.~Ayala.
\newblock The {CAMELS-CL} dataset: catchment attributes and meteorology for
  large sample studies -- {Chile} dataset.
\newblock \emph{Hydrology and Earth System Sciences}, 22\penalty0
  (11):\penalty0 5817--5846, 2018.

\bibitem[Beven(1989)]{Beven1989Physically}
K.~Beven.
\newblock Changing ideas in hydrology --- the case of physically-based models.
\newblock \emph{Journal of Hydrology}, 105\penalty0 (1):\penalty0 157--172,
  1989.
\newblock ISSN 0022-1694.

\bibitem[Beven and Young(2013)]{Beven2013Semantics}
K.~Beven and P.~Young.
\newblock A guide to good practice in modeling semantics for authors and
  referees.
\newblock \emph{Water Resources Research}, 49\penalty0 (8):\penalty0
  5092--5098, 2013.

\bibitem[Beven(2012)]{Beven2012Primer}
K.~J. Beven.
\newblock \emph{Rainfall-runoff modelling: the primer}.
\newblock Wiley-Blackwell, Hoboken, 2nd edition, 2012.
\newblock ISBN 9780470714591.

\bibitem[Coxon et~al.(2020)Coxon, Addor, Bloomfield, Freer, Fry, Hannaford,
  Howden, Lane, Lewis, Robinson, Wagener, and Woods]{Coxon2020CamelsGB}
G.~Coxon, N.~Addor, J.~P. Bloomfield, J.~Freer, M.~Fry, J.~Hannaford, N.~J.~K.
  Howden, R.~Lane, M.~Lewis, E.~L. Robinson, T.~Wagener, and R.~Woods.
\newblock Catchment attributes and hydro-meteorological timeseries for 671
  catchments across {Great Britain} ({CAMELS-GB}), 2020.

\bibitem[Croley(1983)]{Croley1983LBRM}
T.~E. Croley.
\newblock {Great Lake} basins ({U.S.A.-Canada}) runoff modeling.
\newblock \emph{Journal of Hydrology}, 64\penalty0 (1):\penalty0 135--158,
  1983.
\newblock ISSN 0022-1694.

\bibitem[Field et~al.(2012)Field, Barros, Stocker, and Dahe]{IPCC2012}
C.~B. Field, V.~Barros, T.~F. Stocker, and Q.~Dahe.
\newblock \emph{Managing the Risks of Extreme Events and Disasters to Advance
  Climate Change Adaptation: Special Report of the {Intergovernmental Panel on
  Climate Change}}.
\newblock Cambridge University Press, 2012.
\newblock ISBN 9781107025066.

\bibitem[Fiering(1967)]{Fiering1971ABC}
M.~B. Fiering.
\newblock \emph{Streamflow synthesis}, volume~1.
\newblock Harvard University Press, 1967.

\bibitem[Gauch et~al.(2019)Gauch, Mai, and Lin]{Gauch2019Feeding}
M.~Gauch, J.~Mai, and J.~Lin.
\newblock The proper care and feeding of {CAMELS}: How limited training data
  affects streamflow prediction, 2019.
\newblock \textit{arXiv:1911.07249}.

\bibitem[Gupta et~al.(2009)Gupta, Kling, Yilmaz, and Martinez]{Gupta2009NSE}
H.~V. Gupta, H.~Kling, K.~K. Yilmaz, and G.~F. Martinez.
\newblock Decomposition of the mean squared error and {NSE} performance
  criteria: implications for improving hydrological modelling.
\newblock \emph{Journal of Hydrology}, 377\penalty0 (1-2):\penalty0 80--91,
  2009.

\bibitem[Hamman et~al.(2018)Hamman, Nijssen, Bohn, Gergel, and
  Mao]{Hamman2018vic}
J.~J. Hamman, B.~Nijssen, T.~J. Bohn, D.~R. Gergel, and Y.~Mao.
\newblock The {Variable Infiltration Capacity} model version 5 ({VIC-5}):
  infrastructure improvements for new applications and reproducibility.
\newblock \emph{Geoscientific Model Development}, 11\penalty0 (8), 2018.
\newblock ISSN 1991-9603.

\bibitem[Jia et~al.(2020)Jia, Willard, Karpatne, Read, Zwart, Steinbach, and
  Kumar]{Jia2020Physics}
X.~Jia, J.~Willard, A.~Karpatne, J.~S. Read, J.~A. Zwart, M.~Steinbach, and
  V.~Kumar.
\newblock Physics-guided machine learning for scientific discovery: An
  application in simulating lake temperature profiles, 2020.
\newblock \textit{arXiv:2001.11086}.

\bibitem[Karpatne et~al.(2017)Karpatne, Atluri, Faghmous, Steinbach, Banerjee,
  Ganguly, Shekhar, Samatova, and Kumar]{Karpatne2017Theory}
A.~Karpatne, G.~Atluri, J.~H. Faghmous, M.~Steinbach, A.~Banerjee, A.~Ganguly,
  S.~Shekhar, N.~Samatova, and V.~Kumar.
\newblock Theory-guided data science: A new paradigm for scientific discovery
  from data.
\newblock \emph{IEEE Transactions on Knowledge and Data Engineering},
  29\penalty0 (10):\penalty0 2318--2331, 2017.
\newblock ISSN 2326-3865.

\bibitem[Kratzert et~al.(2019)Kratzert, Klotz, Shalev, Klambauer, Hochreiter,
  and Nearing]{Kratzert2019Benchmark}
F.~Kratzert, D.~Klotz, G.~Shalev, G.~Klambauer, S.~Hochreiter, and G.~Nearing.
\newblock Benchmarking a catchment-aware {Long Short-Term Memory} network
  ({LSTM}) for large-scale hydrological modeling.
\newblock \emph{Hydrology and Earth System Sciences Discussions}, pages 1--32,
  2019.

\bibitem[Liang et~al.(1994)Liang, Lettenmaier, Wood, and Burges]{Liang1994vic}
X.~Liang, D.~P. Lettenmaier, E.~F. Wood, and S.~J. Burges.
\newblock A simple hydrologically based model of land surface water and energy
  fluxes for general circulation models.
\newblock \emph{Journal of Geophysical Research}, 99\penalty0 (D7):\penalty0
  14415--14428, 1994.

\bibitem[Newman et~al.(2014)Newman, Sampson, Clark, Bock, Viger, and
  Blodgett]{Newman2014Camels}
A.~Newman, K.~Sampson, M.~P. Clark, A.~Bock, R.~Viger, and D.~Blodgett.
\newblock A large-sample watershed-scale hydrometeorological dataset for the
  contiguous {USA}, 2014.

\bibitem[Trenberth et~al.(2014)Trenberth, Dai, Van Der~Schrier, Jones,
  Barichivich, Briffa, and Sheffield]{Trenberth2014drought}
K.~E. Trenberth, A.~Dai, G.~Van Der~Schrier, P.~D. Jones, J.~Barichivich, K.~R.
  Briffa, and J.~Sheffield.
\newblock Global warming and changes in drought.
\newblock \emph{Nature Climate Change}, 4\penalty0 (1):\penalty0 17--22, 2014.

\end{thebibliography}

\appendix
\section{A Very Simple Process-Based Model}
\label{sec:abcmodel}

\begin{figure}[ht]
    \centering
    \resizebox{0.9\textwidth}{!}{\def\hdist{2}
\def\wdist{10}
\def\cornerrad{2.45}

\begin{tikzpicture}

    \node[font=\large] (groundwater) at (-0.5*\wdist,-\hdist) {Groundwater};
    \node[font=\large] (streamflow)  at (0.25*\wdist,-\hdist) {Streamflow};

    \draw[\niceArrow, rounded corners=\cornerrad] (0,\hdist) -- (0,0) node [pos=0.5,right] {Precipitation};
    \draw[\niceArrow, rounded corners=\cornerrad] (0,0) -- (-0.5*\wdist,0) -- (groundwater) node [pos=0.5,right] {Percolation, $a$};
    \draw[\niceArrow, rounded corners=\cornerrad] (0,0) -- (0.5*\wdist,0) -- (0.5*\wdist,\hdist) node [pos=0.5,right] {Evapotranspiration, $b$};
    \draw[\niceArrow, rounded corners=\cornerrad] (0,0) -- (0.25*\wdist,0) -- (streamflow) node [pos=0.5,right] {$1-a-b$};
    \draw[\niceArrow, rounded corners=\cornerrad] (groundwater) -- (streamflow) node [pos=0.5,above]{Drainage, $c$};

\end{tikzpicture}}
    \caption{Schematic visualization of the ABC-model. The parameters control the fraction $a$ of precipitation that drains into the groundwater storage $G$, the fraction $b$ that evapotranspires, and the fraction $c$ of the groundwater that drains into the stream.}
    \label{fig:abc}
\end{figure}
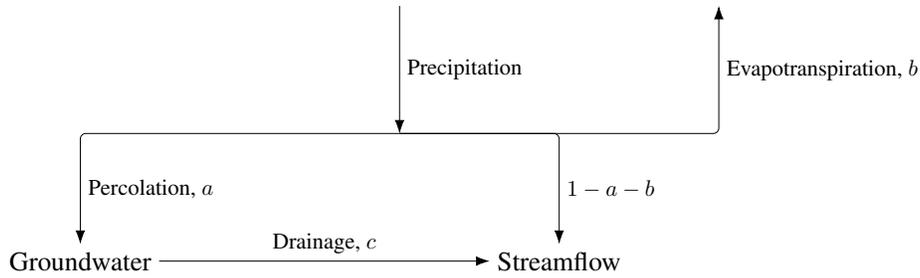
To illustrate the concepts of parameters and storages that process-based models use, we introduce the \textit{ABC-Model}~\cite{Fiering1971ABC}.
It is a simplistic educational model that is not used in practice, but nevertheless helpful to illustrate a lumped process-based model (some would argue that its high level of abstraction makes it not truly process-based but rather purely conceptual, but this discussion is out of scope for this introduction).
The model only takes a one-dimensional precipitation time-series as input.
\autoref{fig:abc} visualizes the schematic setup with three parameters, $a$, $b$, and $c$, and one storage (or model state) $G$.
Parameter $b$ represents \textit{evapotranspiration}, the fraction of rain that evaporates or transpires from the soil, lakes, and plants and consequently never affects streamflow.
$a$ models the fraction of precipitation that trickles through the soil into the groundwater storage $G$.
Finally, $c$ is the fraction of groundwater that flows from the storage into the stream at each time step.

This schematic setup translates into the following equations:
\begin{equation}
\begin{aligned}
    \hat{y}_t &= (1 - a - b) X_t + c G_{t-1} \\
    G_t &= (1 - c) G_{t-1} + a X_t
\end{aligned}
\end{equation}

During model calibration, we search suitable values for the parameters $a, b, c$.
In the warm-up period, we start with some initial value for $G_1$, and generate predictions.
We discard the predictions right away, but keep the updates to $G$.
Finally, we can test the model on the time span that immediately follows the warm-up phase.

\end{document}